\documentclass[12pt]{article}
\usepackage{graphics}
\usepackage{epsfig} 
\usepackage{hyperref}
\usepackage{graphics}
\usepackage{color}      
\usepackage{graphicx}
\usepackage{amsmath, amssymb, graphics}

\usepackage{latexsym,amssymb}

\title{ Exact analytical expressions for  entropy production and  free energy} 

\author{Mesfin Asfaw  Taye$$\thanks{ Electronic address:
mesfin.taye@csun.edu}
\\ 
Department of Physics and Astronomy, California State University\\ Northridge, California, USA }

\begin{document}
\maketitle

\begin{abstract}

The nonequilibrium thermodynamics feature of 
 a Brownian motor operating between two different heat baths is explored as a function of time $t$. Using the Gibbs entropy and 
Schnakenberg microscopic stochastic approach, we find exact closed form expressions for the free energy, the rate of entropy production and the rate of entropy flow from the system to the outside. We show that when the system is out of equilibrium, it constantly produces entropy  
and at the same time extract entropy out of the system. Its entropy production and extraction rates decrease in time and saturate to constant value. In long time limit, the rate of entropy production balances the rate of entropy extraction and at equilibrium both entropy production and extraction rates become zero. Furthermore, via the present model, not only many thermodynamic theories can be   checked but also the wrong conclusions that are given due to lack of exact analytic expressions will be corrected.

\end{abstract}
\maketitle
 

 \section{Introduction}

Equilibrium thermodynamics is well established and intensively studied discipline. However, its application is limited since most systems in nature are far from equilibrium and this warrant the need to advance the nonequilibrium thermodynamics. On other hand,  nonequilibrium thermodynamics deals with systems which are nonhomogeneous where the systems   thermodynamic quantities such as entropy, and free energy strictly rely on the system parameters in complicated manner. As a result nonequilibrium thermodynamics is still under investigation. In the last few decades several studies have been conducted to explore the nonequilibrium features of systems out of equilibrium \cite{mu1,mu2,mu3,mu4}. Particularly 
for system where its dynamics is governed by a master equation,
Boltzmann-Gibbs nonequilibrium entropy
along with the 
entropy balance equation
serves as a basic tool to understand the nonequilibrium thermodynamic features \cite{mu1,mu2,mu3}.

Recent study by Schnakenberg also reveals that various thermodynamics  quantities such as  entropy production rate can be rewritten in terms of local probability density and transition probability rate \cite{mu3}. Based on  this microscopic stochastic approach, many theoretical studies have been conducted \cite{mu4,mu5,mu6,mu7,mu8,mu9,mu10,mu11,mu12,mu13,mu14,mu15,mu16}. These theoretical works confirmed that  systems which are out of equilibrium constantly produce entropy  
and at the same time extract entropy out of the system. In long time limit, the rate of entropy production balances the rate of entropy extraction and at equilibrium both entropy production and extraction rates become zero.    However, most of these previous  studies focused  on exploring the thermodynamic properties of the nonequilibrium system at  steady state \cite{mu1,mu2}.  In  this work we present a simple model which serves as a basic tool  for  a  better understanding  of the nonequilibrium statistical  physics not only in the regime of nonequilibrium steady state (NESS)  but also at any time $t$. We explore the short time behavior of the system either   for isothermal case with load or in general for nonisothermal case with or without load.  Many  mathematical theories can be independently checked via the present model \cite{mu1}. For instance, we show that the entropy balance equation  always satisfied for any parameter choice. Moreover, the first and second laws of thermodynamics are rewritten in terms of the model parameters. Several thermodynamic relations are also uncovered based on the exact analytic results.

At this point we want to stress that in this work, we extend (reconsider) the previous work \cite{mu17} and uncover far more results. 
We find exact closed form expressions for the free energy dissipation rate ${\dot F}$, the rate of entropy production ${\dot e}_{p}$ and the rate of entropy flow from the system to the outside ${\dot h}_{d}$.  The dependence for the change in total entropy, free energy and entropy production on model parameters is studied. Since closed form expressions are obtained for all thermodynamic quantities which are under consideration, we are able to extract thermodynamic information at any time $t$. We believe that even though, a specific model system is considered, the result obtained in this work is generic and advances nonequilibruim thermodynamics.

Moreover it is found that the entropy $S$ attains a zero value at $t=0$; it increases with $t$ and then attains an optimum value. It then decreases as $t$ increases further. In the limit $t\to  \infty$, $S$ approaches a certain constant. Far from equilibrium, in the presence of nonuniform temperature or non-zero load,  ${\dot e}_{p}$ and ${\dot h}_{d}$ decrease with time and  approach their steady state value. At steady state, ${\dot e}_{p}= {\dot h}_{d}>0$. At equilibrium, (for isothermal case and zero load), ${\dot e}_{p}= {\dot h}_{d}=0$.  Moreover, when the heat exchange via kinetic energy is included,  we show that even at quasistatic limit ${\dot e}_{p} \ne 0$ or  ${\dot E}_{p} \ne 0$.  This also implies that since the motor is arranged to undergo a biased random work on spatially arranged thermal background,   Carnot efficiency or Carnot refrigerator  is unattainable;  there is always an irreversible heat flow via the kinetic energy \cite{mu17,mu18,mu19,mu20,mu21}.

The rest of paper is organized as follows: in Section II, we present the model. In Section III,  we derive the expressions the rate of entropy production ${\dot e}_{p}$ and the rate of  entropy flow from the system to the outside ${\dot h}_{d}$.  
In section IV we explore the free energy $F$ as a function of  $t$. The role of particle recrossing  on the entropy production rate will be studied in section V. Section VI deals with summary
and conclusion.

 \begin{figure}[ht]
\centering
{
    \includegraphics[width=6cm]{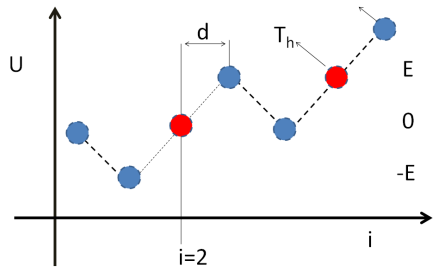}}
\caption{(Color online)  Schematic diagram for a Brownian particle  walking in a discrete ratchet potential with load. 
Sites with red  circles are coupled to the hot reservoir ($T_h$)
while sites with blue circles are coupled to the cold reservoir
($T_c$). Site 2 is labeled explicitly and d is the lattice spacing. } 
\end{figure}

\section{ The model}

Consider a single  Brownian particle that hops  along  one dimensional  discrete ratchet potential   with load  \cite{mu13}
\begin{equation} 
U_i = E[i(mod)3 - 1]+ifd
\end{equation} 
that coupled  with the temperature
\begin{equation} 
  T_i=\{\begin{array}{cl}
   T_{h},&if~ E[i(mod)3 - 1]=0;\\
   T_{c},&otherwise;\end{array}
   \end{equation}
as shown in Fig. 1. The potential  $E>0$, $f$ denotes the load and     $i$ is an integer
that runs from $-\infty$ to $\infty$.  $T_{h}$ and $T_{c}$ denote the temperature for the hot and cold reservoirs, respectively.  Moreover, the one dimensional lattice has  spacing $d$ and in one cycle, the particle walks a net displacement of three  lattice sites.

The jump probability  for the particle to hop from site $i$ to $i+1$ is given by 
$\Gamma e^{-\Delta E/k_{B}T_{i}}$
 where $\Delta E = U_{i+1} - U_i$ and $\Gamma$
is the probability attempting a jump per unit time.  $k_{B}$ designates the Boltzmann constant and  hereafter $k_{B}$, $\Gamma$  and $d$ are  considered  to be a unity.  When the particle undergoes a biased random  walk, it is assumed that first it decides which way to jump (backward or forward)  with equal probability obeying the metropolis algorism. Accordingly, 
when  $\Delta E \le 0$,  the jump
the jump definitely takes place while $\Delta E > 0$  the jump
takes place with probability $\exp(-\Delta E/T_{i})$.

The model exhibits identical behavior with a spin-1 particle system  \cite{mu17}. The master equation which governs the system dynamics is given by
\begin{equation} 
{dP_{n} \over dt}=\sum_{n\neq n'}\left(P_{nn'}p_{n'}-P_{n'n}p_{n}\right),~~n,n'=1,2,3
\end{equation} 
where 
$P_{n'n}$  is the transition probability rate at which the system, originally in state $n$, makes transition to state $n'$.
Here $P_{n'n}$ is given by the Metropolis rule. For instance,
$ 
P_{21}={1 \over 2}e^{-(E+f)/T_{c}}, ~ P_{12}={1 \over 2},~P_{32}={1 \over 2}e^{-(E+f)/T_{h}}$ and $P_{23}={1 \over 2}.
$  The rate equation for the model can then be
expressed as a matrix equation
$
{d \vec{p}  \over dt}= {\bold P}\vec{p}
$ 
where $\vec{p}=(p_{1},p_{2},p_{3})^T$.
${\bold P}$ is a 3 by 3 matrix which
is given by
\begin{equation} 
{\bold P}= \left( \begin{array}{ccc}
{-\mu a^2-\mu^2\over 2a} & {1\over 2} & {1\over 2} \\
{\mu a \over 2} & {-1-\nu b\over 2} & {1\over 2}\\
{\mu^2 \over 2a} & {\nu b\over 2} & -1 \end{array} \right)
\end{equation} 
as long as $0<f<2E/d$.
Here $\mu=e^{-E/T_{c}}$, $\nu=e^{-E/T_{h}}$,  $a = e^{-f d/ T_c}$ and $b = e^{-f d/ T_h}$.
Note that the sum of
each column of the matrix ${\bold P}$  is zero, $\sum_{m}{\bold P}_{mn}=0$ which reveals that  the total probability is conserved: $(d/dt)\sum_{n}p_{n}=d/dt({\bold 1}^T\cdot p)={\bold 1}^T\cdot( {\bold P}\vec{p})=0$.

For the particle which is initially  situated at  site $i=2$,  by solving Eq. (3), we find the time dependent  normalized probability distributions for   $p_{1}(t)$, $p_{2}(t)$  and $p_{3}(t)
$
as shown in Appendix A. The expression for the velocity $V(t)$ at any time $t$ as well as  the  rate of heat flow from the hot reservoir  ${\dot Q}_{h}(t)$ and the rate of heat flow into the cold reservoir  ${\dot Q}_{c}(t)$ are also given in Appendix A.

Hereafter all figures are plotted by taking dimensionless quantities $\epsilon=E/T_{c}$, $\lambda=fd/T_{c}$ and $\tau={T_{h}\over T_{c}}-1$. We also introduce dimensionless time ${\bar t}=\Gamma t$ and after this the bar will be dropped.

\section{Derivation of Entropy and entropy production}

 The relation between the internal energy $U(t)$, entropy $S(t)$ and free  energy $F(t)$   is well known for isothermal system (uniform temperature $T$) which satisfies detailed balance condition,   and can be written as
\begin{eqnarray}
{\dot U}[p_{i}(t)]&=&-T{\dot h}_{d}(t)=-{\dot H}_{d}(t)\\ \nonumber
&=&-T\sum_{i>j}(p_{i}P_{ji}-p_{j}P_{ij}) \ln \left({P_{ji}\over P_{ij}}\right),
\end{eqnarray}
\begin{eqnarray}
{\dot F}[p_{i}(t)]&=&-T{\dot e}_{p}(t)=-{\dot E}_{P}(t)\\ \nonumber
&=&-T\sum_{i>j}(p_{i}P_{ji}-p_{j}P_{ij}) \ln \left({p_{i}P_{ji}\over p_{j} P_{ij}}\right)
\end{eqnarray}
and the corresponding fundamental entropy balance equation is given by
\begin{eqnarray}
{\dot S(t)}&=&e_{p}(t)-h_{d}(t)
\end{eqnarray}
where $S$ is 
the Gibbs entropy  given by  
\begin{eqnarray}
S[{p_{i}(t)}]&=&-\sum_{i=1}^N p_{i} \ln p_{i}.
\end{eqnarray}
Here ${\dot h}_{d}(t)$  and ${\dot e}_{p}(t)$ designate  the term which is related to the heat dissipation rate  and the instantaneous entropy production (${\dot e}_{p}>0$), respectively. The entropy balance equation (7) can be rewritten as ${\dot S}^{T}(t)={\dot E}_{p}(t)-{\dot H}_{d}(t)$ where $S^T(t)={TdS(t)/ dt}$.
Recently  for isothermal system which is driven out of equilibrium, the above relations have been extended via phonological approach \cite{mu1}.

Next, we  examine   whether the well-known thermodynamic relations, which are valid for  equilibrium system,  are still obeyed  for  the model  system which is driven out of equilibrium due to inhomogeneous thermal arrangement $T_{h} \ne T_{c}$ or non-zero external load $f \ne 0$. 
  For the three state model we present,     
 \begin{eqnarray}
S[{p_{i}(t)}]=-\sum_{i=1}^3 p_{i} \ln p_{i}, 
\end{eqnarray}
\begin{eqnarray}
{\dot h}_{d}&=&\sum_{i>j}(p_{i}P_{ji}-p_{j}P_{ij}) \ln \left({P_{ji}\over P_{ij}}\right)\nonumber \\
&=&(-p_{1}P_{21}+p_{2}P_{12})\ln\left({P_{12}\over P_{21}}\right) +\nonumber \\
&&(-p_{2}P_{32}+p_{3}P_{23})\ln\left({P_{23}\over P_{32}}\right)+ \nonumber \\
&&(p_{3}P_{13}-p_{1}P_{31})\ln\left({P_{13}\over P_{31}}\right)
\end{eqnarray}
and
\begin{eqnarray}
{\dot e}_{p}&=&\sum_{i>j}(p_{i}P_{ji}-p_{j}P_{ij}) \ln \left({p_{i}P_{ji}\over p_{j} P_{ij}}\right)\nonumber \\
&=&(-p_{1}P_{21}+p_{2}P_{12})\ln\left({p_{2}P_{12}\over p_{1}P_{21}}\right) +\nonumber \\
&&(-p_{2}P_{32}+p_{3}P_{23})\ln\left({p_{3}P_{23}\over p_{2}P_{32}}\right)+ \nonumber \\
&&(p_{3}P_{13}-p_{1}P_{31})\ln\left({p_{3}P_{13}\over p_{1}P_{31}}\right)
\end{eqnarray}
where the indexes $i=1... 3$ and $j=1...3$. 
It is important to note that via the expressions   $p_{1}(t)$, $p_{2}(t)$  and $p_{3}(t) $ that are shown in Appendix A and using the rates (see Eq. (4)),
\begin{eqnarray}
P_{21}&=&{1 \over 2}e^{-(E+f)/T_{c}} ~P_{12}={1 \over 2},~P_{32}={1 \over 2}e^{-(E+f)/T_{h}}\nonumber \\
 && P_{23}={1 \over 2}, P_{13}={1 \over 2},  P_{31}={1 \over 2} e^{-(2E-f)/T_{c}} 
\end{eqnarray}
 the thermodynamic quantities which are under investigation can be evaluated.

{ \it For our nonequilibrium system where $T_{h} \ne T_{c}$ and $f \ne 0$,  regardless  of  our parameter choice, we find that 
the fundamental entropy balance equation 
\begin{eqnarray}
{\dot S}(t)&=&{\dot e}_{p}(t)-{\dot h}_{d}(t)
\end{eqnarray}
is always  satisfied  at any time $t$}. 
Moreover  ${\dot h_{d}}(t)$ is found to satisfy the relation
\begin{eqnarray}
{\dot h}_{d}(t)&=&{-{\dot Q}_{h}(t) \over T_{h}}+{{\dot Q}_{c}(t) \over T_{c}}.
\end{eqnarray}
Next we study how $S$, ${\dot S}$, and ${\dot e}_{p}$ and $  {\dot h}_{d}$  vary in time.

{\it Entropy.\textemdash}  The  entropy of the system  exhibits an intriguing parameter dependence. 
 Exploiting Eq. (9) one can see that  
(for any parameter choice) when $t \to 0$, $S  \to 0$. As time increases $S$ increases and attains an optimum value. As $t$ further increases, $S$ decreases.   In the limit  $ t \to \infty$,  the total entropy approach  its steady state  value $S \to  -\sum_{i=1}^3 p_{i}^s \ln p_{i}^s$.  Particularly, for the case where $f=0$, one can get a simplified expression by expanding  $S$ in small $t$ regime and  after some algebra one gets 
\begin{eqnarray}
S&=&{1\over 2}e^{{-U_{0}\over T_{h}}} (1+e^{{U_{0}\over T_{h}}}+e^{{U_{0}\over T_{h}}}\ln[2] )t   \\  \nonumber
 && -{1\over 2}e^{{-U_{0}\over T_{h}}}(\ln[{e^{{-U_{0}\over T_{h}}}\over 2}]- \ln[t]-e^{{U_{0}\over T_{h}}} \ln[t] )t.
\end{eqnarray}
Eq. (15) clearly depicts that in the limit  $t \to 0$, $S  \to 0$.
In Fig. 2a, we plot the entropy $S$ as a function of $t$  for parameter choice of $U_0=2.0$ , $\tau=2$  and $\lambda=0$ (red line)  and   $\lambda=0.5$ (black line). The figure once again shows that the entropy   
attains an optimum value at particular time $t$.

 One can note that the system sustains a non-vanishing current (the system is out of equilibrium) as long as $T_{h} \ne T_{c}$ and   $U_{0}\ne 0$ both in the presence or in the absence of load. For isothermal case, the system is driven  out of equilibrium  only in the presence of load    $f> 0$. This implies that  
the system relaxes to equilibrium in the absence of load and when  $T_{h} = T_{c}$.  In the  absence of both  external load and bistable  potential $U_0=0$, the particle undergoes a random walk on lattice. In long time limit, the system may  approach equilibrium as long as the heat exchange via kinetic energy is neglected  (even if a distinct temperature difference is retained between the hot and cold reservoirs). Thus when $f=0$ and in the limit $U_0 \to 0$ (approaching equilibrium), Eq. (9) converges to   
\begin{eqnarray}
 S(t)&=&{1\over 3}e^{{-3t\over 2}}(-2(-1+e^{{3t\over 2}})\ln{({1\over 3}-{e^{{-3t\over 2}}\over 3})})  \nonumber \\
&&-{1\over 3}e^{{-3t\over 2}}((2+e^{{3t\over 2}})\ln{({1\over 3}+{2\over 3}e^{{-3t\over 2}})}).
\end{eqnarray}
As it can be seen from Eq. (16),  in the limit $S\to 0$ as $t\to 0$ and when $t\to \infty$, $S\to \ln[3]$.   Note that  at stationary state, $S$ is constant  implies that $\Delta S=0$.
\begin{figure}[ht]
\centering
{
    \includegraphics[width=6cm]{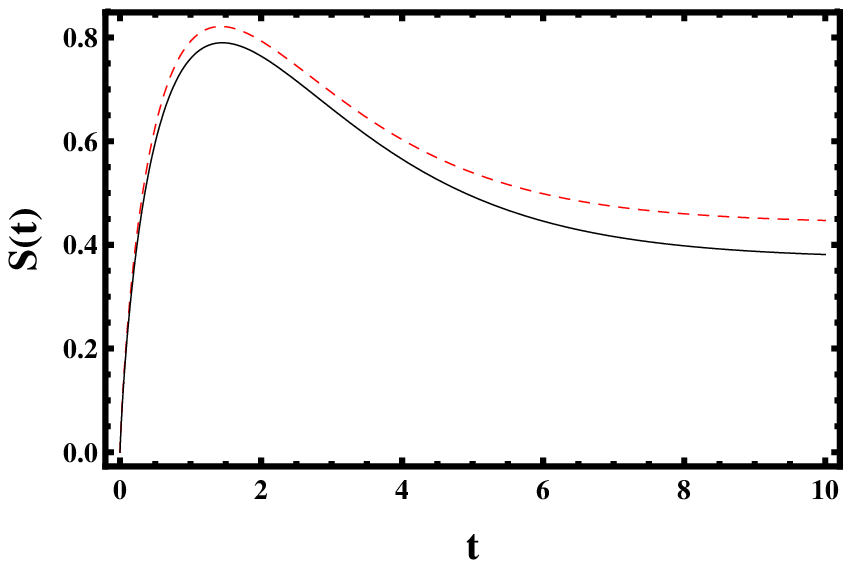}}
\hspace{1cm}
{
    \includegraphics[width=6cm]{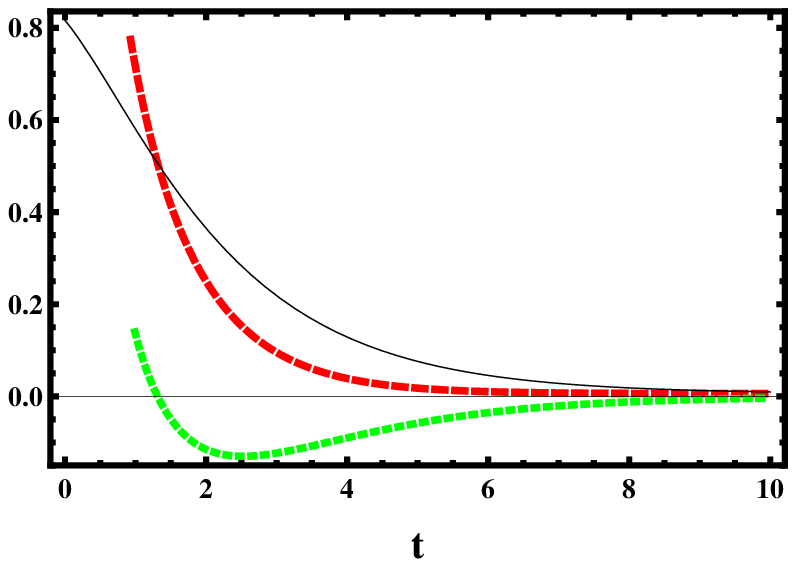}
}
\caption{ (Color online) (a) Total entropy $S(t)$  versus $t$ for a given  $\epsilon=2$,  $\tau=2.0$ and $\lambda=0$.(b) Plot of ${\dot S}(t)$ (dotted green  line), ${\dot e}_{p}(t)$ (dashed red line) and ${\dot h}_{d}(t)$ (black solid line) as a function of  $t$  for a given  $\epsilon=2$,  $\tau=2.0$ and $\lambda=0$.     } 
\label{fig:sub} 
\end{figure}
\begin{figure}[ht]
\centering
{
    \includegraphics[width=6cm]{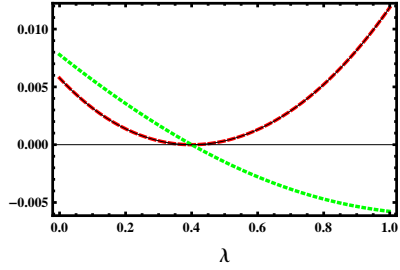}
}
\caption{ (Color online) (a) Plot of ${\dot S}(t)$ (green line), ${\dot e}_{p}(t)$ (dashed red line) and ${\dot h}_{d}(t)$ (solid line) as a function of  $t$  for a given  $\epsilon=2$,  $\tau=2.0$ and $t=10^6$ (steady state).  Here  ${\dot e}_{p}(t)={\dot h}_{d}(t)$. } 
\label{fig:sub} 
\end{figure}

{\it Entropy production rate.\textemdash} Let us focus on the rate of entropy production ${\dot e}_{p}$, the rate of total entropy  ${\dot S}(t)$ and the rate of entropy flow from the system to the outside  ${\dot h}_{d}(t)$. The plot of ${\dot S}(t)$ (dotted green line), ${\dot e}_{p}(t)$ (dashed red line) and ${\dot h}_{d}(t)$ (solid black line) as a function of  $t$  is depicted in Fig. 2b  for a given values of  $\epsilon=2$,  $\tau=2.0$ and $\lambda=0$. 
 The  figure indicates that ${\dot e}_{p}$ and ${\dot h}_{d}(t)$ decrease   to their steady state values  ${\dot e}_{p}= {\dot h}_{d}>0.04$ as time progress. 
  Exploiting Eqs. (10), (11) and (13), one can also see that far from equilibrium  ${\dot e}_{p}>0$ and ${\dot h}_{d}>0$. When $t \to 0$, ${\dot e}_{p}$ becomes much greater than ${\dot h}_{d}(t)$ and  as time increases, in a certain time interval, ${\dot h}_{d}(t)>{\dot e}_{p}$.  
	In general, as time progresses
  ${\dot e}_{p}$ and ${\dot h}_{d}$ decrease and approach their steady state value. At steady state, ${\dot e}_{p}= {\dot h}_{d}>0$. At equilibrium, for isothermal case and zero load, ${\dot e}_{p}= {\dot h}_{d}=0$. Particularly expanding ${\dot e}_{p}$ and ${\dot h}_{d}(t)$  in the small time regime, we find 
	\begin{eqnarray}
 {\dot e}_{p}={1\over 2}e^{{-U_{0}\over T_{h}}}(\ln[{2\over t}]-e^{{-U_{0}\over T_{h}}}\ln[{e^{{-U_{0}\over T_{c}}}\over 2}]-(e^{{U_{0}\over T_{h}}})\ln[t])
\end{eqnarray}
and
\begin{eqnarray}
 {\dot h}_{d}={-1\over 2}\ln[e^{{-U_{0}\over T_{c}}}]-{1\over 2}e^{{-U_{0}\over T_{h}}}\ln[{e^{{U_{0}\over T_{h}}}}].
\end{eqnarray}
 On the other hand in the limit
 $U_{0} \to 0$ and $f=0$ (approaching equilibrium),  ${\dot h}_{d}=0$  for any $t$ while 
\begin{eqnarray}
 {\dot e}_{p}=-e^{{-3t\over 2}}\ln[{-1+e^{{3t\over 2}}\over 2+e^{{3t\over 2}}}].
\end{eqnarray}
Here  ${\dot e}_{p}>0$ for small $t$ and decreases  (the system relaxes to equilibrium) as time increases. 
In the limit $t \to \infty$, ${\dot e}_{p}={\dot h}_{d}=0$.

The  quasistatic limit of the system corresponds  to the case where the velocity  approaches  zero  $ V(t)=0$ in the vicinity of the stall force or in the absence of load when $U_0 \to 0$.   The general expression for the stall force is lengthy and will not be presented.  However, at steady state, the stall force reduces to  
$
f={E ({T_{h}\over T_{c}}-1)/ (2 {T_{h}\over T_{c}}+1)}
$. At quasistatic limit,  regardless of any parameter choice, we find ${\dot e}_{p}={\dot h}_{d}(t)=0$. This can be appreciated by plotting ${\dot e}_{p}$ or ${\dot h}_{d}(t)$ as a function of load  in long time limit.  Figure 3  exhibits that both ${\dot e}_{p}$ and ${\dot h}_{d}(t)$ decrease as the load increases and attain a zero value at stall force  $\lambda=0.4$. As the load further increases,  
${\dot e}_{p}$and ${\dot h}_{d}(t)$ step up. On the contrary,  ${\dot S}$ decreases with time. At  stall force, ${\dot S}=0$.
\begin{figure}[ht]
\centering
{
    \includegraphics[width=6cm]{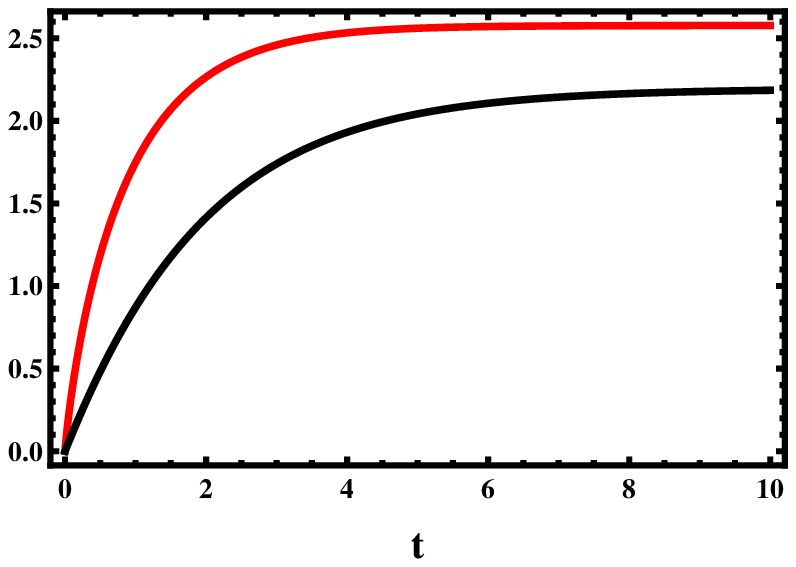}}
\hspace{1cm}
{
    \includegraphics[width=6cm]{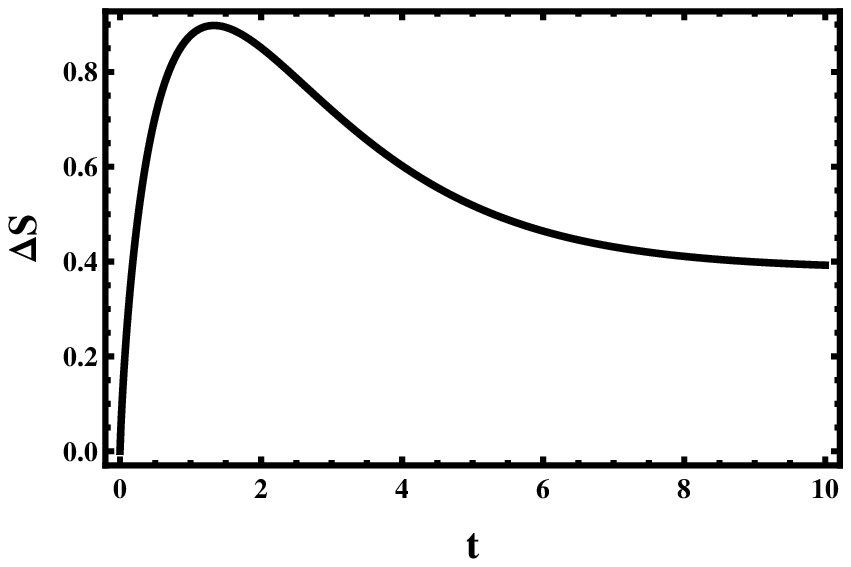}
}
\caption{ (Color online) (a) Plot of   $\Delta e_p=e_p(t)-e_p(0)$ (red line) and ${\Delta  h}_{d}=h_d(t)-h_d(0)$ (Black line) as a function of  $t$  for a given  $\epsilon=2$,  $\tau=2.0$ and $\lambda=0$.  (b)  $\Delta S(t)=S(t)-S(0)$  versus $t$ for a given  $\epsilon=2$,  $\tau=2.0$ and $\lambda=0$.   } 
\label{fig:sub} 
\end{figure}

Since   closed form expressions for 
${\dot S}(t)$, ${\dot e}_{p}(t)$  and ${\dot h}_{d}(t)$ are obtained for any time $t$, the analytic expressions for  the change in entropy production,  heat dissipation  and  the change in  the total entropy can be found analytically via   
\begin{eqnarray}
\Delta h_d&=& \int_{t_0}^{t}\left(\sum_{i>j}(p_{i}P_{ji}-p_{j}P_{ij}) \ln \left({P_{ji}\over P_{ij}}\right)\right)dt,
\end{eqnarray}
 \begin{eqnarray}
\Delta e_{p}&=& \int_{t_0}^{t} \left(\sum_{i>j}(p_{i}P_{ji}-p_{j}P_{ij}) \ln \left({p_{i}P_{ji}\over p_{j}P_{ij}}\right)\right)dt 
\end{eqnarray}
and 
\begin{eqnarray}
\Delta S &=&\int_{t_0}^{t} \left(\sum_{i>j}(p_{i}P_{ji}-p_{j}P_{ij}) \ln \left({p_{i}\over p_{j}}\right)\right)dt \nonumber \\
&=&-\sum_{i=1}^3 p_{i}(t) \ln p_{i}(t)+\sum_{i=1}^3 p_{i}(0) \ln p_{i}(0)
\end{eqnarray}
where $\Delta S=\Delta e_p-\Delta h_d$ and the indexes $i=1... 3$ and $j=1...3$.  
The expressions for $\Delta h_d$,  $\Delta S$  and $\Delta e_p$  are lengthy.

Here care must be taken since  ${\dot e}_{p}(t)>0$  or  ${\dot h}_{d}(t)>0$ does not imply that $\Delta h_d=h_d(t)-h_d(t_0)>0$,  $\Delta S=S(t)-S(t_0)>0$  or $\Delta e_p=e_p(t)-e_p(t_0)>0$. Rather since ${\dot h}_d$,  ${\dot S}$  and  ${\dot e}_p$ are decreasing function of $t$,  $\Delta h_d$,  $\Delta S$  and $\Delta e_p$ can take even negative values  depending on the interval between $t_0$ and $t$. If the system reaches at steady state or stationary state and if the change in these parameters are taken at  the steady state regime, then $\Delta h_d=0$,  $\Delta S=0$  or $\Delta e_p=0$. However this does not indicate that second law of thermodynamics is violated.

In fact the second law of thermodynamics is always preserved  as it can be seen by exploiting the exact analytic expressions (Eqs. (20-22).   
In reality, once the motor starts operating, entropy will be accumulated in the system starting from $t=0$ and as time progresses, more entropy will be  stored in the system even though some entropy is extracted out of the system. Hence 
if the change in these parameters is taken between $t=0$ and any time $t$, always the inequality 
$\Delta h_d=h_d(t)-h_d(0)>0$,  $\Delta S=S(t)-S(0)>0$  or $\Delta e_p=e_p(t)-e_p(0)>0$ holds true and as time progresses the change in this parameters increases as shown in Fig. 4a. 
As a matter of fact,  in small $t$ regimes,  ${\dot e}_{p}$ becomes much greater than ${\dot h}_{d}(t)$ (see Fig. 2b) revealing that the entropy production is higher (than entropy extraction)  in  the first few period of times.  As time increases, more entropy will be  extracted   ${\dot h}_{d}(t)>{\dot e}_{p}$. Over all, since the system produces enormous amount of  entropy in initial times,  in latter time or any time $t$, $\Delta e_p>\Delta h _d$ and hence $\Delta S>0$ (see Fig. 4).

\section{Free energy dissipation rate and Free energy}

In order to relate the free energy dissipation rate with ${\dot E}_{p}(t)$   and  ${\dot H}_{d}(t)$ let us 
now introduce  ${\dot H}_{d}(t)$ for the model system we considered. The heat dissipation rate is given by ${\dot H}_{d}(t)=-{\dot Q}_{h}(t)+{\dot Q}_{c}(t)$. Using Eqs. (56) and (57), one finds  
\begin{eqnarray}
{\dot H}_{d}(t)&=&-{\dot Q}_{h}(t)+{\dot Q}_{c}(t)\nonumber \\
&=& \sum_{i>j}T_{j}(p_{i}P_{ji}-p_{j}P_{ij}) \ln \left({P_{ji}\over P_{ij}}\right). 
\end{eqnarray}
Equation (23) is notably different from Eq. (10) due to 
the term  $T_{j}$. Similar relation has been used by Hao. $et$. $al$  for the isothermal case \cite{mu1}. We rewrite Eq. (23) as 
\begin{eqnarray}
{\dot H}_{d}(t)&=& \sum_{i>j}T_{j}(p_{i}P_{ji}-p_{j}P_{ij}) \ln \left({P_{ji}\over P_{ij}}\right)  \nonumber \\
&=&\sum_{i>j}T_{j}(p_{i}P_{ji}-p_{j}P_{ij}) \ln \left({p_{i}P_{ji}\over p_{j}P_{ij}}\right)-\nonumber \\
&&\sum_{i>j}T_{j}(p_{i}P_{ji}-p_{j}P_{ij}) \ln \left({p_{i}\over p_{j}}\right) \nonumber \\
&=& {\dot E}_{p}(t)-{\dot S}^T(t)
\end{eqnarray}
where 
 \begin{eqnarray}
{\dot E}_{p}(t)&=& \sum_{i>j}T_{j}(p_{i}P_{ji}-p_{j}P_{ij}) \ln \left({p_{i}P_{ji}\over p_{j}P_{ij}}\right)
\end{eqnarray}
and 
\begin{eqnarray}
{\dot S}^T(t)&=&\sum_{i>j}T_{j}(p_{i}P_{ji}-p_{j}P_{ij}) \ln \left({p_{i}\over p_{j}}\right).
\end{eqnarray}
Here the indexes $i=1... 3$ and $j=1...3$.
Now we have entropy balance equation  ${\dot S}^T(t)={\dot E}_{p}(t)-{\dot H}_{d}(t)$  for our model system.  
Note that for isothermal case  Eqs. (24), (25) and (26)
 converge to  $T {\dot h}_{d}(t)$, $T {\dot e}_{p}(t)$ and 
$
{\dot S}^{T}(t)=T {\dot S}(t)
$,
respectively.  All the above analysis  indicates that ${\dot E}_{p}(t)$, $ {\dot H}_{d}(t)$ and  ${\dot S}^T(t)$  contain  a term $(p_{2}P_{32}-p_{3}P_{23})$  which is  associated  with the rate of heat taken out of the hot reservoir and two terms $(p_{2}P_{12}-p_{1}P_{21})$ and $(p_{3}P_{13}-p_{1}P_{31})$ which are associated with  the rate of heat given to the cold reservoir.

\begin{figure}[ht]
\centering
{
    \includegraphics[width=6cm]{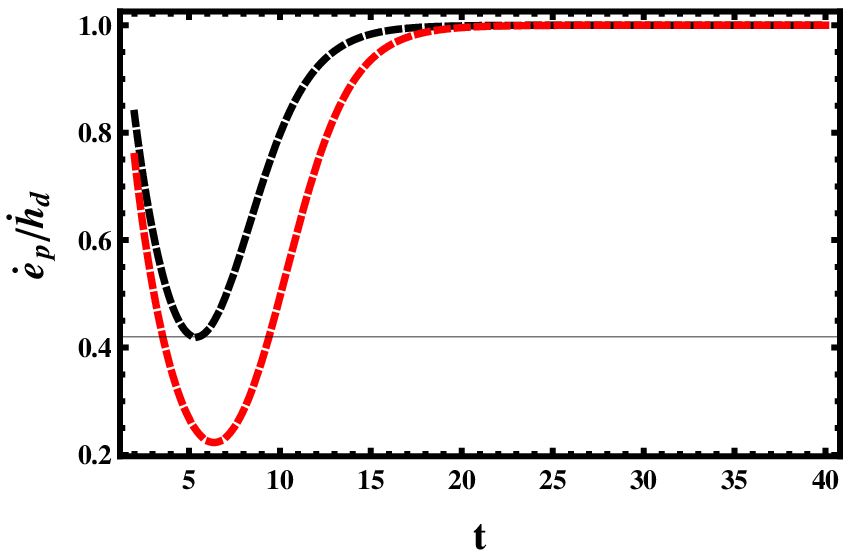}}
\hspace{1cm}
{
    \includegraphics[width=6cm]{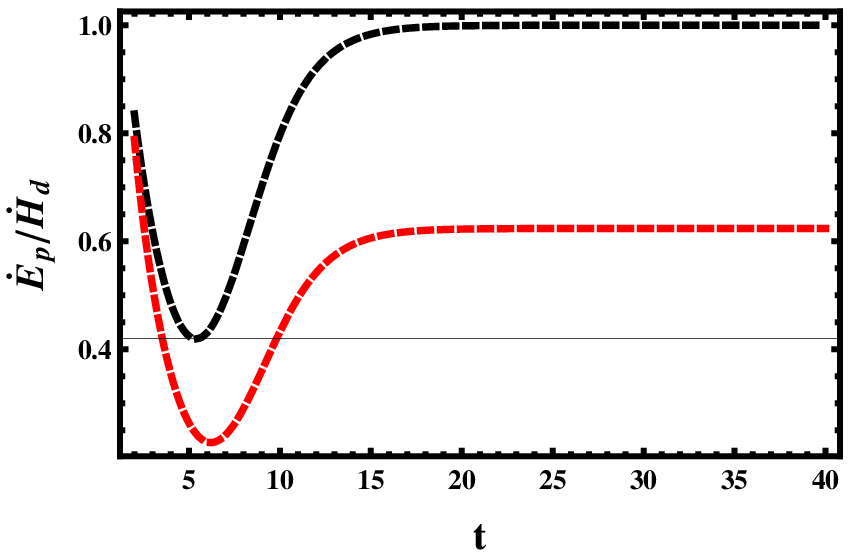}
}
\caption{ (Color online) (a) Plot of  ${\dot e}_{p}(t)/ {\dot h}_{d}(t)$  as a function of  $t$  for a given  $\epsilon=2$, $f=0.8$  $\tau=2.0$ (red line) and $\tau=1.0$ (black line).  (b) (a) Plot of  ${\dot E}_{p}(t)/ {\dot H}_{d}(t)$  as a function of  $t$  for a given  $\epsilon=2$, $f=0.8$  $\tau=2.0$ (red line) and $\tau=1.0$ (black line).     } 
\label{fig:sub} 
\end{figure}

\begin{figure}[ht]
\centering
{
    \includegraphics[width=6cm]{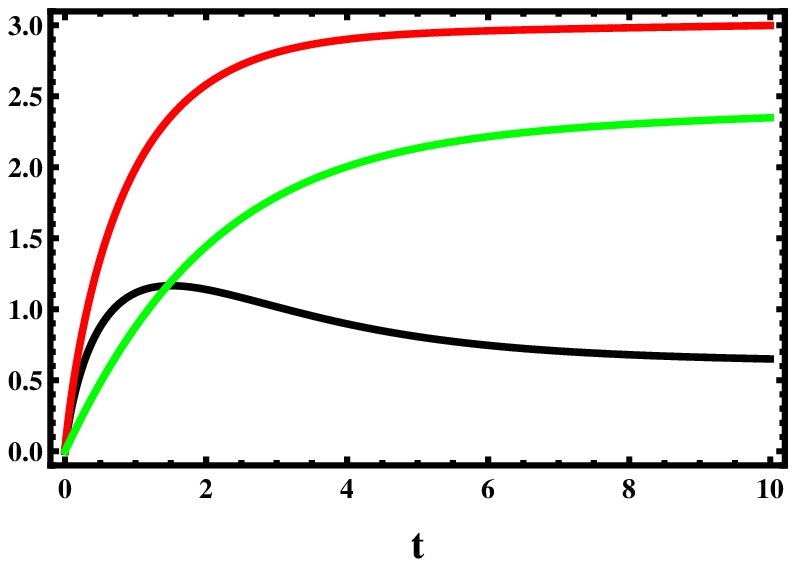}}
\hspace{1cm}
{
    \includegraphics[width=6cm]{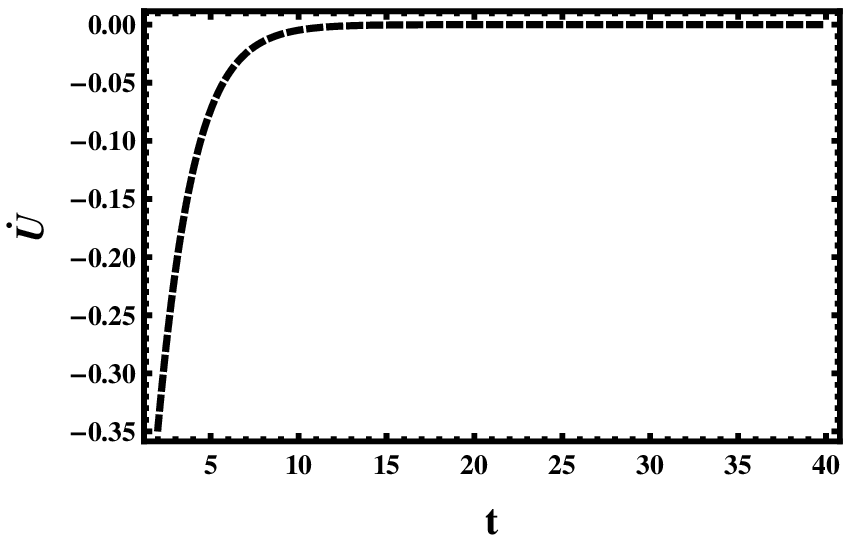}}

\caption{ (Color online) (a) Plot of  ${\Delta E}_{p}(t$ (red line)  $\Delta H_{d}(t)$ (green line) and  ${\Delta S}^T(t)$ (black line) as a function of  $t$  for a given  $\epsilon=2$, $f=0.8$ and  $\tau=2.0$ (red line).  (b)  Plot of  ${\dot U}(t)$  as a function of  $t$  for a given  the same parameter choice.     } 
\label{fig:sub} 
\end{figure}

\begin{figure}[ht]
\centering
{
    \includegraphics[width=6cm]{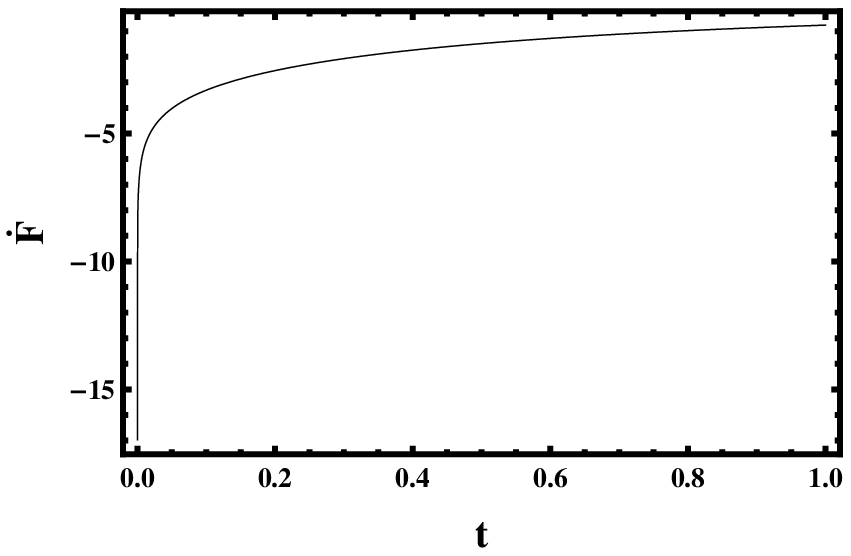}}
\hspace{1cm}
{
    \includegraphics[width=6cm]{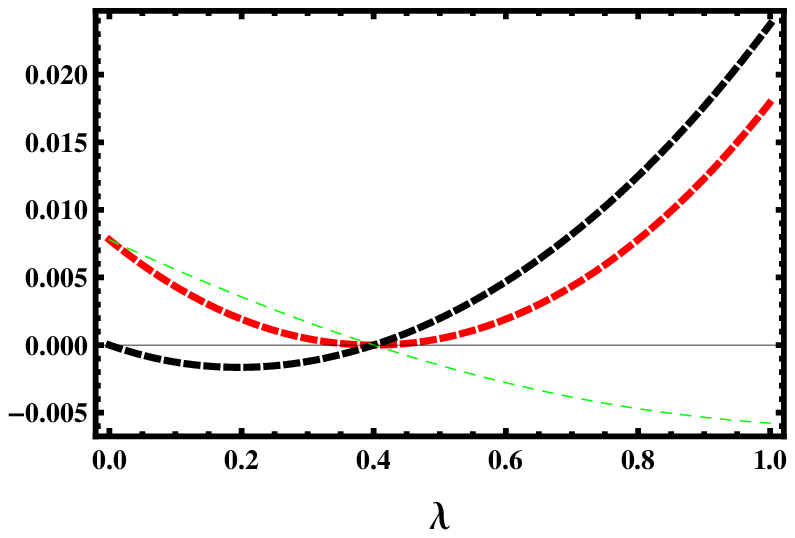}}
\caption{ (Color online) (a) The free energy dissipation rate  ${\dot F}$  versus $t$ for a given  $\epsilon=2$,  $\tau=2.0$ and $\lambda=0.8$.  (b) Plot of ${\dot S^T}(t)$ (dotted line), ${\dot E}_{p}(t)$ (dashed line) and ${\dot H}_{d}(t)$ (solid line) as a function of  $f$  for a given  $\epsilon=2$,  $\tau=2.0$ and $\lambda=0$.   } 
\label{fig:sub} 
\end{figure}
One notable difference between  ${\dot e}_{p}(t) / {\dot h}_{d}(t)$ and   ${\dot E}_{p}(t) / {\dot H}_{d}(t)$  is that ${\dot e}_{p}(t) / {\dot h}_{d}(t)$ always approaches  unity in long time limit while ${\dot E}_{p}(t) / {\dot H}_{d}(t)$   approaches unity only for isothermal cases with or without load as shown in  Fig. 5. Figure 5a depicts the  plot of  ${\dot e}_{p}(t)/ {\dot h}_{d}(t)$  as a function of  $t$  for a given  $\epsilon=2$, $f=0.8$  $\tau=2.0$ (red line) and $\tau=1.0$ (black line).  Figure 5b  shows the Plot of  ${\dot E}_{p}(t)/ {\dot H}_{d}(t)$  as a function of  $t$  for a given  $\epsilon=2$, $f=0.8$ and  $\tau=2.0$ (red line) and $\tau=1.0$ (black line).

The second law of thermodynamics  can be   written  $\Delta S^T=\Delta  E_{p}-\Delta H_{d}$ where $\Delta S^T$, $\Delta  E_{p}$ and $\Delta H_{d}$ are very lengthy expressions which can be evaluated  via 
\begin{eqnarray}
\Delta H_d&=& \int_{t_0}^{t}\left(\sum_{i>j}T_{j}(p_{i}P_{ji}-p_{j}P_{ij}) \ln \left({P_{ji}\over P_{ij}}\right)\right)dt,
\end{eqnarray}
 \begin{eqnarray}
\Delta E_{p}&=& \int_{t_0}^{t} \left(\sum_{i>j}T_{j}(p_{i}P_{ji}-p_{j}P_{ij}) \ln \left({p_{i}P_{ji}\over p_{j}P_{ij}}\right)\right)dt
\end{eqnarray}
and 
\begin{eqnarray}
\Delta S^T &=&\int_{t_0}^{t} \left(\sum_{i>j}T_{j}(p_{i}P_{ji}-p_{j}P_{ij}) \ln \left({p_{i}\over p_{j}}\right)\right)dt.
\end{eqnarray}
In Fig. 6a, we plot ${\Delta E}_{p}=E_p(t)-E_p(0)$ (red line), $\Delta H_{d}(t)=H_d(t)-H_d(0)$ (green line) and  ${\Delta S}^T(t)=S^{T}(t)-S^{T}(0)$ (black line) as a function of  $t$  for given  $\epsilon=2$, $f=0.8$ and  $\tau=2.0$. The figure exhibits that as long as  $t>0$,   $\Delta E_p>\Delta H _d$ and hence $\Delta S^T>0$.

On the other hand, the total internal energy $U(t)$ is the sum of the internal energies \cite{mu23}
\begin{eqnarray}
U[{p_{i}(t)}]&=&\sum_{i=1}^3 p_{i}u_{i} \nonumber \\
&=&p_{1}(t)(-E)+p_{3}(t)(E)
\end{eqnarray}
while the change in the internal energy is given by 
\begin{eqnarray}
\Delta U&=&U[{p_{i}(t)}]-U[{p_{i}(0)}]\nonumber \\
&=&E\left(p_{3}(t)-p_{3}(0)+p_{1}(0)-p_{1}(t)\right).
\end{eqnarray}

As the particle walks along the reaction coordinate, it receives some heat from the hot reservoir and gives part of it to the cold bath. The remaining heat will be spent to do some work ${\dot W(t)}=fV(t)$ against the external load. Hence we  verify  the first law  of thermodynamics 
\begin{eqnarray}
{\dot U}[P_{i}(t)]&=&-\sum_{i>j}(p_{i}P_{ji}-p_{j}P_{ij}) \left(u_{i}-u_{j}\right) \nonumber \\
&=&{\dot Q}_{h}(t)-{\dot Q}_{c}(t)-{\dot W}(t)\nonumber \\
&=&-({\dot H}_{d}(t)+fV(t)).
\end{eqnarray}
The change in the internal energy (Eq. (31)) can be also rewritten as 
\begin{eqnarray}
\Delta U&=&-\int_{0}^{t} \left({\dot Q}_{h}(t)-{\dot Q}_{c}(t)-W(t)  \right)dt  \nonumber \\
        &=&  \int_{0}^{t} \left(  -({\dot H}_{d}(t)+fV(t))  \right)dt   \nonumber \\
				&=& \Delta Q_h-\Delta Q_{c}-\Delta W.
\end{eqnarray}
Note that similar relation is derived  in the work \cite{mu23}  for the isothermal case.  Fig. 6b  depicts the  plot of  ${\dot U}(t)$  as a function of  $t$  for the same parameter choice  of   $\epsilon=2$, $f=0.8$ and  $\tau=2.0$. The figure shows that as $t\to \infty$, ${\dot U}(t )\to 0$  as expected.

Next let us find   the expression for the free energy dissipation rate  ${\dot F}$.  For the isothermal case, the free energy is given by $F=U-TS$ and  we can still adapt this relationship to nonisothermal case to write 
\begin{eqnarray}
{\dot F} (t)&=&{\dot U}-{\dot S}^T(t).
\end{eqnarray}  
Substituting Eqs. (23) and (25) in Eq. (32) leads to
\begin{eqnarray}
{\dot F} (t)+{\dot E}_{p}(t)={\dot U }(t)+{\dot H}_{d}(t)=-fV(t)
\end{eqnarray}
which  is the second law of thermodynamics. Note that  
in the absence of load,   ${\dot U }(t)=-{\dot H}_{d}(t)$ and  consequently 
$
{\dot E}_{p}(t)=-{\dot F} (t)
$.   
The dependence of the free energy on time $t$ can be explored   by exploiting Eq. (35). In general ${\dot F}<0$ and approaches to zero in the long time limit (see Fig. 7a). In order to get a very simplified expression, let us expand Eq. (35) in small time regime for the case where $f=0$ to get 
\begin{eqnarray}
 {\dot F}&=&{1\over 2}e^{{-U_{0}\over T_{h}}}(-T_{h}\ln[2]+e^{{U_{0}\over T_{h}}}\ln[{e^{{-U_{0}\over T
_{c}}}\over 2}]T_{c})+\nonumber \\
 &&{1\over 2}e^{{-U_{0}\over T_{h}}}(e^{{U_{0}\over T_{h}}}T_{c}\ln[t]+T_{h}\ln[t]).
\end{eqnarray}
When $t\to 0$, ${\dot F}$ diverges. On the other hand when 
 $U_{0} \to 0$ and $f=0$ (approaching equilibrium),  we get 
\begin{eqnarray}
 {\dot F}=e^{{-3t\over 2}}(T_{c}+T_h)\ln[{e^{{3t\over 2}}-1\over 2+ e^{{3t\over 2}}}].
\end{eqnarray}
As it can been seen in Eq. (37), in the limit $t \to \infty$, ${\dot F} \to 0$  and when  $t \to 0$, ${\dot F}$  diverges.

The change in the free energy also is given by  
\begin{eqnarray}
\Delta F&=&-\int_{t_0}^{t} \left(  fV(t)+ {\dot E}_{p}(t)   \right)dt \\ \nonumber
        &=&  \int_{t_0}^{t} \left(  {\dot U }(t)+{\dot H}_{d}(t)- {\dot E}_{p}(t)   \right)dt\\ \nonumber
				&=& \Delta U+\Delta H_{d}-\Delta E_{p}.
\end{eqnarray}

At  quasistatic limit where the velocity  approaches  zero  $ V(t)=0$, ${\dot E}_{p}(t) =0$ and ${\dot H}_{d}(t) =0$ (see Fig.(7b)) and far from quasistatic limit 
$E_{p}>0$  which is  expected as   the engine operates irreversibly. Far from  stall force,   ${\dot E}_{p}(t) \ne {\dot H}_{d}(t)$ as long as a distinct temperature difference between the hot reservoirs is retained.  However, for isothermal case, at steady state, ${\dot E}_{p}(t) ={\dot H}_{d}(t)$ which implies $\Delta F=\Delta U$.

Finally, in the work \cite{mu1}, the  second law of thermodynamics is rewritten in terms of the "housekeeping heat" and excess heat by considering  isothermal case.   For the model system we consider, when  the particle  undergoes a cyclic motion, at least it has to get $fV(t)$ amount of energy rate from the hot reservoir in order to keep the system at steady state. Hence $ fV(t)$ is equivalent to the  "housekeeping heat" $Q_{hk}$ and we can rewrite Eq. (35)  as 
\begin{eqnarray}
{\dot F} (t)+{\dot E}_{p}(t)={\dot U }(t)+{\dot H}_{d}(t)=-fV(t)=-{\dot Q}_{hk}
\end{eqnarray}
while the excess heat ${\dot Q}_{ex}$ part is given by
\begin{eqnarray}
Q_{ex}={\dot H}_{d}-{\dot Q}_{hk}.
\end{eqnarray}

For isothermal case, we can rewrite the second law of thermodynamics as \begin{eqnarray}
{\dot S}^T={\dot E}_{p}-{\dot H}_{d}={-\dot F}-{\dot Q}_{ex}
\end{eqnarray}
and 
\begin{eqnarray}
{\dot F}={\dot Q}_{hk}-{\dot E}_{p}.
\end{eqnarray}

\section{Irreversibility due to particle recrossing along the boundaries}

In the previous section we have seen that at quasistatic limit, the system is reversible  and hence ${\dot e}_{p}=0$ or  ${\dot E}_{p}=0$ even if $T_{h} \ne T_{c}$. However so far the rate of heat loss due to particle recrossing at the boundary between the hot and cold reservoirs is not included. If the heat exchange via kinetic energy is included,  even at quasistatic limit ${\dot e}_{p} \ne 0$ or  ${\dot E}_{p} \ne 0$.  This also implies that since the motor is arranged to undergo a biased random work on spatially arranged thermal background,   Carnot efficiency or Carnot refrigerator  is unattainable;  there is always an irreversible heat flow via the kinetic energy \cite{mu17,mu18,mu19,mu20,mu21}.
\begin{figure}[ht]
\centering
{
    \includegraphics[width=6cm]{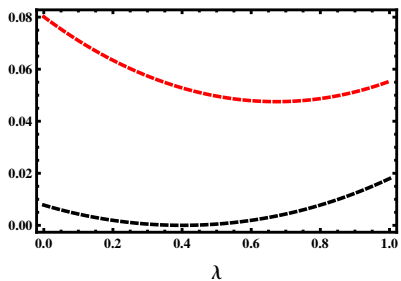}}
\caption{ (Color online)  Plot of ${\dot E}_{p}(t)$ (dashed black line) and ${\dot E}_{p}^{*}(t)$ (red line) as a function of  $\lambda$  for a given  $\epsilon=2$,  $\tau=2.0$ and $t=10^{6}$.    } 
\label{fig:sub} 
\end{figure}

When the particle by chance jumps from the cold to hot reservoir, it absorbs, $k_B(T_h - T_c)/2$ amount of heat from the hot bath. When it hops back to the cold reservoir, it dumps back this $k_B(T_h - T_c)/2$ amount of heat to the cold bath. This irreversible heat flow is always one way, i.e.; from the hot to the cold baths and it cannot be recovered. 
Hence  we can conclude this heat transfer  from the hot to cold reservoirs depends on how often the particle  jumps from the cold to hot heat baths and it can be written as  
\begin{eqnarray}
{\dot Q}_{irr}=\left (p_{1}P_{21}+p_{3}P_{23}\right)(T_{h}-T_{c})/2.
\end{eqnarray}
The heat exchange via kinetic energy does not affect ${\dot H} _{d}$ since the whole heat taken from the hot reservoirs goes to the cold reservoir. This also implies that the whole heat dumped to the cold reservoirs contributes to  the internal entropy production   and hence for any parameter choice  ${\dot E}_{p}\ge 0$ as long as $T_{h} \ne T_{c}$.
{\it The heat loss due to particle recrossing  only contributes to the internal entropy production and we infer the new entropy production rate to be}
\begin{eqnarray}
{\dot E}_{p}^{*} ={\dot E}_{p} +{\dot Q}_{irr}.
\end{eqnarray}
One can then rewrite the thermodynamic relations that derived in the previous section in terms of 
${\dot E}_{p}^{*}$. Here ${\dot E}_{p}^{*}=0$ only when  $T_{h} = T_{c}$. ${\dot E}_{p}^{*}$ as a function of the load $\lambda$ is plotted in Fig. 8.  The figure depicts that even at quasistatic limit ${\dot e}_{p}^{*}\ne 0$. The same figure shows that if the heat exchange via kinetic energy is omitted, ${\dot E}_{p}=0$ at stall force $\lambda =0.4$. Note that when $U_0$  and $f=0$,  the particle undergoes unbiased random walk problem.  Its average velocity is zero but its speed is nonzero. If the particle by chance hops from the cold to  hot reservoirs, it absorbs an energy and latter dumps this energy to cold bath. So even in this case the system is far from equilibrium.

\section{summary and conclusion}

In this work we present a paradigmatic model  which serves a basic tool  for  a  better understanding  of the nonequilibrium statistical  physics not only in the regime of nonequilibrium steady state (NESS)  but also at any time $t$. Not only the long time behavior of the system is explored but also  we investigate the short time behavior of the system either   for isothermal case with load or in general for nonisothermal case with or without load.

 Based on the Gibbs entropy and 
Schnakenberg microscopic stochastic approach, we get exact analytic expressions for the thermodynamic quantities that are under investigations. We show that whenever the system is 
  out of equilibrium, it  constantly produce entropy  
and at the same time extract entropy out of the system. At steady state, the rate of entropy production balances the rate of entropy extraction and at equilibrium both entropy production and extraction rates become zero.   Moreover we show that the entropy balance equation \cite{mu22}
 always satisfied for any parameter choice. Furthermore, the first and second laws of thermodynamics are rewritten in terms of the model parameters. Several thermodynamic relations are also uncovered based on the exact analytic results.

The exact analytic expressions are also  obtained 
for the free energy dissipation rate ${\dot F}$, the rate of entropy production ${\dot e}_{p}$ and the rate of  entropy flow from the system to the outside ${\dot h}_{d}$.  Particularly, the analytical study reveals that the entropy $S$ attains a zero value at $t=0$; it increases with $t$  and then attains an optimum value. It then decreases as $t$ increases further. In the limit $t \to \infty$, $S$ approaches a certain constant. Far from equilibrium, in the presence of nonuniform temperature or non-zero load,  ${\dot e}_{p}$ and ${\dot h}_{d}$ decrease with time and  approach their steady state value. At steady state, ${\dot e}_{p}= {\dot h}_{d}>0$. At equilibrium, for isothermal case and zero load, ${\dot e}_{p}= {\dot h}_{d}=0$.  Moreover, when the heat exchange via kinetic energy is included,  we show that even at quasistatic limit ${\dot e}_{p} \ne 0$ or  ${\dot E}_{p} \ne 0$.  This also implies that since the motor is arranged to undergo a biased random work on spatially arranged thermal background,   Carnot efficiency or Carnot refrigerator  is unattainable;  there is always an irreversible heat flow via the kinetic energy.

In conclusion, in this work we present  a  simple model which  serves as a basic tool  for   better understanding  of the nonequilibrium statistical  physics  not only in the regime of nonequilibrium steady state (NESS)  but also at any time $t$. The present model also serves as a tool to check many elegant thermodynamic theories. Based on this exactly solvable model, we uncovered several thermodynamic relations. 
We believe that even though, a specific model system is considered, the result obtained in this work is generic and advances nonequilibruim thermodynamics.

\section*{Acknowledgment}
I would like also to thank Mulu  Zebene for her
constant encouragement.

\section*{ Appendix A}
In this Appendix we will give the expressions for $p_{1}(t)$, $p_{2}(t)$  and $p_{3}(t)$ as well as $V(t)$, ${\dot Q}_{h}(t)$ and ${\dot Q}_{c}(t)$
For the particle which is initially  situated at  site $i=2$,  the time dependent  normalized probability distributions after solving Eq. (3) are
\begin{eqnarray}
p_{1}(t)&=&c_{1}\frac{a (2+\nu b)}{\mu \left(\mu+\left(a^2+\mu\right) \nu b\right)}+\\ \nonumber
& &c_{2} e^{-\frac{\left(a+a^2 \mu+\mu^2\right) t}{2 a}} \left(-1+\frac{a
(-1+a \mu)}{-\mu^2+a \nu b}\right),\\
p_{2}(t)&=&-c_{3} e^{\frac{1}{2} t (-2-\nu b)}-c_{2}\frac{a\text{  }e^{-\frac{\left(a+a^2 \mu+\mu^2\right) t}{2 a}} (-1+a \mu)}{-\mu^2+a
\nu b}+\\ \nonumber
&&c_{1}\frac{ \left(2 a^2+\mu\right)}{\mu+\left(a^2+\mu\right) \nu b},\\
p_{3}(t)&=&c_{1}+c_{2} e^{-\frac{\left(a+a^2 \mu+\mu^2\right) t}{2 a}}+
c_{3} e^{\frac{1}{2} t (-2-\nu b)}
\end{eqnarray}
where 
\begin{eqnarray}
c_{1}&=& \frac{\mu \left(\mu+\left(a^2+\mu\right) \nu b\right)}{\left(a+a^2 \mu+\mu^2\right) (2+\nu b)},\\
c_{2}&=& -\frac{a}{\left(a+a^2 \mu+\mu^2\right) \left(-1+\frac{a (-1+a \mu)}{-\mu^2+a \nu b}\right)},\\
c_{3}&=& -\frac{\mu \left(\mu+a^2 \nu b+\mu \nu b\right)}{\left(a+a^2 \mu+\mu^2\right) (2+\nu b)}+ \\ \nonumber
&&\frac{a}{\left(a+a^2 \mu+\mu^2\right) \left(-1+\frac{a (-1+a \mu)}{-\mu^2+a
\nu b}\right)}.
\end{eqnarray}
The  $\sum_{i=1}^3 p_{i}(t)=1$ revealing the probability distribution is normalized.  In the limit of $t \to \infty$, we recapture the steady state probability distributions
\begin{eqnarray}
p_{1}^{s}&=&\frac{a}{a+a^2 \mu +\mu ^2},\\
p_{2}^{s}&=&\frac{\mu  \left(2 a^2+\mu \right)}{\left(a+a^2 \mu +\mu^2\right) (2+b \nu )},\\
p_{3}^{s}&=&\frac{\mu  \left(\mu +b \left(a^2+\mu \right) \nu \right)}{\left(a+a^2 \mu +\mu ^2\right) (2+b \nu)}
\end{eqnarray}
shown in the work \cite{mu13}.

The velocity  $V(t)$ at any time $t$ is the difference between the forward $V_{i}^{+}(t)$ and backward $V_{i}^{-}(t)$ velocities at each site $i$ 
\begin{eqnarray}
V(t)&=& \sum_{i=1}^{3}(V_{i}^{+}(t)-V_{i}^{-}(t)) \\ \nonumber
&=&(p_{1}P_{21}-p_{2}P_{12})+(p_{2}P_{32}-p_{3}P_{23})+\\ \nonumber
&&(p_{3}P_{13}-p_{1}P_{31}).
\end{eqnarray}
Exploiting Eq. (14), one can see that the particle retains a unidirectional current when  $f=0$ and  $T_{h}>T_{c}$.  For isothermal case $T_{h}=T_{c}$, 
   the system sustains a non-zero velocity  in the presence of load $f \ne 0$ as expected. Moreover, when  $t \to \infty$,  
the velocity  $V(t)$ increases with $t$ and approaches to steady state velocity 
\begin{eqnarray}
V^{s}=3{\mu \left(b a \nu-{\mu\over a}\right) \over 2(2+\nu b)\left(1+a\mu+{\mu^2\over a}\right)}
 \end{eqnarray}

The  heat per unit time taken from the hot reservoir has a form 
\begin{eqnarray}
{\dot Q}_{h}(t)&=&(E+f)(p_{2}P_{32}-p_{3}P_{23}) \nonumber \\
               &=&T_{h} (p_{2}P_{32}-p_{3}P_{23})\ln({P_{32}\over P_{23}}).
\end{eqnarray}
On the other hand, the  heat per unit time given to cold reservoir is given by
\begin{eqnarray}
{\dot Q}_{c}(t)&=&(E+f)(p_{2}P_{12}-p_{1}P_{21})+\nonumber \\
&&(2E-f)(p_{3}P_{13}-p_{1}P_{31}) \nonumber \\
&=&T_{c}(p_{2}P_{12}-p_{1}P_{21})\ln({P_{12}\over P_{21}})+ \nonumber \\
&&T_{c}(p_{3}P_{13}-p_{1}P_{31})\ln({P_{13}\over P_{31}})
\end{eqnarray}
As  $t \to \infty$,  both ${\dot Q}_{h}(t)$ and ${\dot Q}_{c}(t)$ evolve in time to  their corresponding  steady state values  
\begin{eqnarray}
{\dot Q}_{h}^s=(E+f){\mu \left(b a \nu-{\mu\over a}\right) \over 2(2+\nu b)\left(1+a\mu+{\mu^2\over a}\right)}
\end{eqnarray}
and
\begin{eqnarray}
{\dot Q}_{c}^s&=&(E-2f){ \mu \left(b a \nu-{\mu\over a}\right) \over 2(2+\nu b)\left(1+a\mu+{\mu^2\over a}\right)}.
\end{eqnarray}

\end{document}